\documentclass[reprint,amsmath,amssymb,prapplied,aps,superscriptaddress,nofootinbib]{revtex4-2}

\usepackage{lipsum}
\usepackage{graphicx}
\usepackage{dcolumn}
\usepackage{bm}
\usepackage[separate-uncertainty=true]{siunitx}
\usepackage[dvipsnames]{xcolor}
\usepackage{hyperref}
\hypersetup{colorlinks,allcolors=MidnightBlue}
\usepackage{url}
\usepackage{float}

\begin{document}

\title{Niobium Air Bridges as Low-Loss Components for Superconducting Quantum Hardware}

\author{N.~Bruckmoser}
\thanks{These authors contributed equally to this work:\\ \href{mailto:Niklas.Bruckmoser@wmi.badw.de}{Niklas.Bruckmoser@wmi.badw.de}, \href{mailto:Leon.Koch@wmi.badw.de}{Leon.Koch@wmi.badw.de}}
\author{L.~Koch} 
\thanks{These authors contributed equally to this work:\\ \href{mailto:Niklas.Bruckmoser@wmi.badw.de}{Niklas.Bruckmoser@wmi.badw.de}, \href{mailto:Leon.Koch@wmi.badw.de}{Leon.Koch@wmi.badw.de}}

\author{I.~Tsitsilin}
\author{M.~Grammer}
\author{D.~Bunch}
\author{L.~Richard}
\author{J.~Schirk}
\author{F.~Wallner}
\author{J.~Feigl}
\author{C.M.F.~Schneider}
\affiliation{Technical University of Munich, TUM School of Natural Sciences, Department of Physics, 85748 Garching, Germany}
\affiliation{Walther-Meißner-Institut, Bayerische Akademie der Wissenschaften, 85748 Garching, Germany}

\author{S.~Gepr{\"a}gs}
\author{V.P.~Bader}
\affiliation{Walther-Meißner-Institut, Bayerische Akademie der Wissenschaften, 85748 Garching, Germany}

\author{M.~Althammer}
\author{L.~S\"odergren}
\affiliation{Technical University of Munich, TUM School of Natural Sciences, Department of Physics, 85748 Garching, Germany}
\affiliation{Walther-Meißner-Institut, Bayerische Akademie der Wissenschaften, 85748 Garching, Germany}
\author{S.~Filipp}
\affiliation{Technical University of Munich, TUM School of Natural Sciences, Department of Physics, 85748 Garching, Germany}
\affiliation{Walther-Meißner-Institut, Bayerische Akademie der Wissenschaften, 85748 Garching, Germany}
\affiliation{Munich Center for Quantum Science and Technology (MCQST), Schellingstra\ss{}e 4, 80799 M\"unchen, Germany}

\date{\today}

\begin{abstract}

Scaling up superconducting quantum processors requires a high routing density for readout and control lines, relying on low-loss interconnects to maintain design flexibility and device performance. We propose and demonstrate a universal subtractive fabrication process for air bridges based on an aluminum hard mask and niobium as the superconducting film. Using this technology, we fabricate superconducting CPW resonators incorporating multiple niobium air bridges in and across their center conductors. Through rigorous cleaning methods, we achieve mean internal quality factors in the single-photon limit exceeding  $Q_{\mathrm{int}} = 8.2 \times 10^6$. Notably, the loss per air bridge remains below the detection threshold of the resonators. Due to the larger superconducting energy gap of niobium compared to conventional aluminum air bridges, our approach enables stable performance at elevated temperatures and magnetic fields, which we experimentally confirm in temperatures up to 3.9 K and in a magnetic field of up to 1.60 T. Additionally, we utilize air bridges to realize low-loss vacuum-gap capacitors and demonstrate their successful integration into transmon qubits by achieving median qubit lifetimes of  $T_1 = \SI{51.6}{\micro s}$.

\end{abstract}

\maketitle

\section{\label{sec:intro}Introduction}

In quantum as well as classical microwave frequency integrated circuits, air bridges are an integral element to route signals by leaving the substrate and thus acting as a crossover between perpendicular paths.
These interconnects are necessary to enable large-scale quantum processors by increasing the density and complexity of coplanar waveguide (CPW) routing \cite{krinner.2022, valles-sanclemente2023}. In addition, they play an important role in mitigating unwanted crosstalk between adjacent signal lines and parasitic modes \cite{Simons.2001, ponchak.2005, Arute.2019, cao.2023}.
Due to their complexity, many different fabrication processes have been proposed based on both optical \cite{Schicke.2003, chen2014, dunsworth.2018, sun2022, bu.2025, alegria.2025} and electron beam lithography \cite{ girgis2006, khalid2012, jin2021, janzen2022, stavenga2023, valles-sanclemente2023}. 
With a few exceptions \cite{Schicke.2003, stavenga2023, bu.2025, alegria.2025, tao.2024}, most superconducting processes employ aluminum (Al) for air bridge metallization, typically fabricating these structures after the Josephson junctions \cite{Wu.2021, valles-sanclemente2023}. However, this sequencing imposes strict limitations: aggressive cleaning methods such as hydrofluoric acid or Piranha solution cannot be used to remove surface oxides, as they would damage Al-based structures, including the junctions themselves. Additionally, post-junction processing steps must adhere to a minimal thermal budget to preserve junction integrity \cite{korshakov.2024}. By instead fabricating air bridges before Josephson junctions, higher processing temperatures - during baking, sputtering, and etching - can be applied without compromising junction performance. This approach enables the use of temperature-sensitive techniques, such as laser annealing, to fine-tune qubit frequencies and mitigate frequency crowding \cite{hertzberg2021, zhang.2022}.
Furthermore, the superconducting energy gap of niobium (Nb) is significantly higher than Al \cite{Finnemore.1966}. This leads to a higher resilience of superconductivity in the presence of magnetic fields and thus also allows for galvanically connected Nb air bridges to be active elements within flux lines with typical currents in the order of $\SI{1}{mA}$ \cite{krinner2019}. 
In addition, the higher superconducting critical temperature enables quantum experiments at elevated temperatures above \SI{200}{mK} \cite{anferov.2020, anferov.2024}.\\
\indent A key difficulty in using Nb for air bridges is that soft resist masks pose processing challenges. In particular, sputtered atoms penetrate underlying layers due to its high kinetic energy. \cite{chandrasekaran.2021, murthy.2022}. This leads to lossy interfaces on the bottom side of the air bridges due to the high amount of two-level systems (TLS) that are hosted in photoresist \cite{Quintana.2014}.
To improve on these aspects, we develop a fabrication process based on a sacrificial hard mask that enables any complementary superconducting material such as Nb, tantalum (Ta) or niobium-titanium nitride (NbTiN) to be used as an air bridge. 
Due to its low etch resistance, we use Al as a sacrificial hard mask, which allows us to easily remove the mask afterwards. 
With this method we achieve quality factors substantially higher compared to existing fabrication processes \cite{chen2014,bu.2025, tao.2024}, and well comparable to standard non-airbridge processes. These can therefore be readily used as low-loss routing elements. Moreover, our low-loss air bridge technology can also be used to form a transmon qubit by creating a capacitive shunt between the ground plane and a large overlying air bridge.

\begin{figure*}
\includegraphics[width=0.98\textwidth]{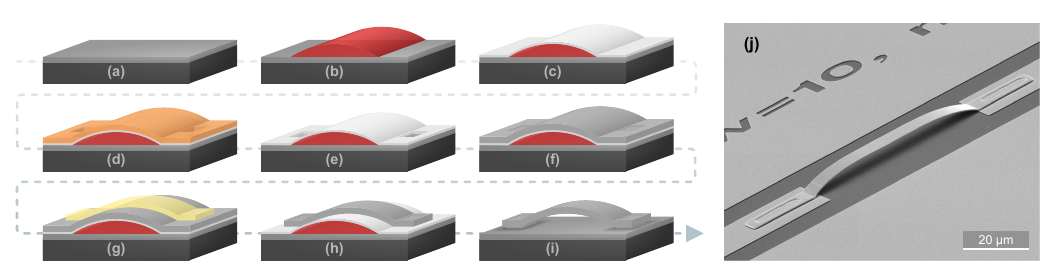}
\caption{\label{fig:fig1}Fabrication steps of superconducting niobium air bridges. (a)\,Metallization and ground plane patterning with reactive ion etching, (b)\,optical lithography and reflow of positive resist, (c)\, angled evaporation of a sacrificial Al hard mask, (d)\,contact area opening by first defining a thin optical resist mask, followed by wet etching of the hard mask, (e)\,Selective resist removal by oxygen ashing, (f)\,removal of surface oxides on the Nb groundplane via argon (Ar) ion milling and sputtering of a thick Nb air bridge layer, (g)\,patterning of optical resist to define lateral bridge dimensions, (h)\,pattern transfer with SF$_6$-based reactive ion etching, (i)\,Al hard mask removal by wet chemical etching and resist stripping with an NMP based remover. (j)\,Scanning electron microscope (SEM) image of an air bridge with a length of \SI{60}{\micro m} and a width of \SI{10}{\micro m}.}
\end{figure*}

\section{\label{sec:fabrication}Fabrication}

In order to realize low-loss Nb sputtered air bridges, we utilize an additional thin hard mask on top of the frequently used soft mask, illustrated in Fig.\,\ref{fig:fig1}\,(c) as part of the fabrication process flow. The hard mask serves three essential purposes in our fabrication process. First, it acts as a protective barrier for the bottom resist, mitigating ion implantation during argon ion milling \cite{okuyama.1978}. Second, it functions as a shielding layer during sputtering, preventing intermixing between the underlying resist and high kinetic Nb particles \cite{chandrasekaran.2021, murthy.2022}. Finally, the hard mask serves as a dedicated stopping layer in the reactive ion etching process \cite{tian.2008}. This ensures precise control to prevent etching into the Nb ground plane.

Prior to the air bridge processing steps, we fabricate the groundplane of the chips containing macroscopic ($\geq~\SI{2}{\micro m}$) structures such as coplanar waveguides, ports, and qubit pads.
We start with high-resistivity ($>~\SI{10}{k\ohm cm}$) \SI{100}{mm} silicon (Si) wafers that are cleaned in Piranha acid (3 parts \SI{96}{wt\%} $\mathrm{H_2SO_4}$ to 1 part \SI{30}{wt\%} $\mathrm{H_2O_2}$) and a 7:1 buffered oxide etch (BOE) solution (7 parts \SI{40}{wt\%} $\mathrm{NH_4F}$ to 1 part \SI{49}{wt\%} $\mathrm{HF}$) to remove lossy silicon oxides \cite{Wisbey.2010}. Afterwards, we immediately transfer the substrate in an ultra-high vacuum system (PLASSYS MEB550 S4-I) to sputter-deposit a \SI{150}{nm} thin film of Nb. Using maskless optical lithography, we pattern coplanar waveguide structures and transfer these into the film using an $\mathrm{SF}_6$-based reactive ion etching process. After a wet-chemical resist stripping step (TechniStrip P1331), these samples act as the starting point of the hard mask based air bridge process [cf.\ Fig.\,\ref{fig:fig1}\,(a)-(i)].
 
We control the bridge height of $\SI{2.7}{\micro m}$ by spin coating the reflow-compatible resist AZ ECI 3027 on the patterned wafer. The reflow step at \SI{140}{\degree C} for \SI{5}{minutes} is needed to reduce the steepness of sidewalls and ensure stability of released air bridges [Fig.\,\ref{fig:fig1}\,(b)]. We afterwards evaporate a thin sacrificial layer of \SI{100}{nm} Al as a hard mask at a wafer tilt of \SI{20}{\degree} to cover topographies in the ground plane [Fig.\,\ref{fig:fig1}\,(c)]. This hard mask needs to be sufficiently thick to serve as a reproducible etch stop layer during the reactive ion etching process of the air bridges. We deposit the hard mask with electron beam evaporation instead of sputtering in order to reduce intermixing in the resist-aluminum interface. To create contact areas to the underlying Nb layer, we pattern a thin optical resist (AZ MiR 701) on the pads of the air bridge [Fig.\,\ref{fig:fig1}\,(d)]. Due to its alkalinity, the TMAH based developer directly transfers the exposed pattern into the hard mask. We subsequently use an oxygen plasma to selectively remove the top resist without attacking the hard mask and reflowed bottom resist [Fig.\,\ref{fig:fig1}\,(e)]. Afterwards, we load the wafer in a sputtering chamber and use argon ion milling to remove surface oxides on the underlying Nb groundplane. This step is essential to ensure galvanic connection between the air bridge and the groundplane. We then sputter-deposit a \SI{900}{\nano m} thick layer of Nb. We use a low pressure of \SI{1.3}{mTorr} to reduce intrinsic stress in the film \cite{imamura1991} [Fig.\,\ref{fig:fig1}\,(f)]. In our case, optimization of this step has been found to be essential in order to ensure high stability of fabricated air bridges. High tensile stress leads to breaking points between the pad and the arch, which is further described in Appendix \ref{sec:supp-stress}.  Afterwards, we coat the niobium layer with the etch-resistant negative photoresist ma-N 1420 to define width and length of the air bridge [Fig.\,\ref{fig:fig1}\,(g)]. We use an $\mathrm{SF}_6$ based process to etch the top niobium layer in multiple in-situ steps to reduce thermal load and thus prevent reflow of the underlying bottom resist [Fig.\,\ref{fig:fig1}\,(h)]. The Al hard mask serves as an etch stopper due to its resilience against $\mathrm{SF}_6$-based etching chemistry. Afterwards, we remove the hard mask with a wet-chemical Al etchant (TechniEtch Al80) at an elevated temperature of \SI{60}{\degree C} and use an NMP based remover to strip the underlying resist [Fig.\,\ref{fig:fig1}\,(i)]. The air bridges are dried by nitrogen blow-drying after an immersion in hot isopropyl alcohol (IPA) due to its low surface tension coefficient \cite{bolgar.2025}. We avoid using sonication for cleaning, as it may result in rip-offs at the contact areas.\\
\indent The process as such is compatible as an "air bridge last" process, i.e.\ air bridges are fabricated as the last element of a device, since sensitive elements like Josephson junctions are protected by resist during etching. However, by fabricating air bridges prior to Al based elements, we are able to employ additional rigorous cleaning steps. We remove potential crosslinked resist and hard mask residues by immersing the chip for 10 minutes in Piranha acid, followed by a 10 minute BOE immersion to remove lossy niobium oxides \cite{Altoe.2020}. Afterwards, we immerse the wafer a second time for 10 minutes in Piranha and 20 minutes in BOE to remove buried defects and residues in the metal-air interface.
A fully processed air bridge is shown in Fig.\,\ref{fig:fig1}\,(j). The air bridge has a width of \SI{10}{\micro m} and a length of \SI{60}{\micro m}.
Bridges up to a length of \SI{60}{\micro m} show a yield of \SI{100}{\percent} for approximately 23500 fabricated air bridges on a four inch wafer.

\section{\label{sec:air-bridge-characterization}air bridge characterization}

\begin{figure}
\includegraphics{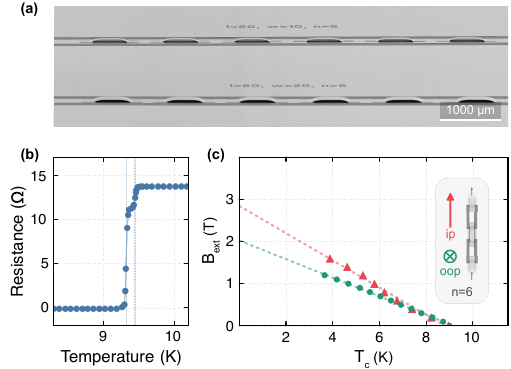}

\caption{\label{fig:transport}Transport measurements of multiple air bridges chained together. \textbf{(a)} SEM image of a daisy chain with $n_\mathrm{AB}=6$ air bridges. \textbf{(b)} Resistance vs.\ temperature for the daisy chain at zero magnetic field. The first transition at $T_\mathrm{c,AB}=\SI{9.32}{K}$ corresponds to the air bridge, the second transition at $T_\mathrm{c,film}=\SI{9.43}{K}$ corresponds to the thin film ground plane. \textbf{(c)} Magnetic field $B_\mathrm{c}$ for in-plane (red) and out-of-plane geometry (green).}
\end{figure}

In order to ensure galvanic connection and to investigate superconducting properties, we measure resistance as a function of temperature across a daisy chain of six air bridges [cf.\ Fig.\,\ref{fig:transport}\,(a)]. Each air bridge has a width of $\SI{10}{\micro m}$ and a length of $\SI{60}{\micro m}$. The corresponding resistance curve is shown in Fig.\,\ref{fig:transport}\,(b). By fitting two Gaussians to the derivative of the resistance, we identify two superconducting transitions at $T_\mathrm{c,AB}=\SI{9.32}{K}$ and $T_\mathrm{c,film}=\SI{9.43}{K}$. The value for the air bridge transition $T_\mathrm{c,AB}$ is close to the transition temperature found in literature \cite{Rairden.1964} and thus indicates that sufficiently high quality Nb films can be grown on Al seed layers. Due to its higher critical temperature compared to Al, Nb air bridges can be used in superconducting integrated circuits operating well above $\SI{1}{K}$.

The high $T_\mathrm{c}$ also relates to a high upper critical magnetic field $B_\mathrm{c2,0}$. We extract the critical temperature by applying a constant external magnetic field $B_\mathrm{ext}$ and sweeping the temperature $T$. By repeating this measurement for different values of $B_\mathrm{ext}$, we can extrapolate the critical field $B_\mathrm{c2,0}$ at zero temperature by fitting the measured data points to the Ginzburg-Landau theory with $B_\mathrm{c2}(T) = B_\mathrm{c2,0}(1-T/T_\mathrm{c})$. From Fig.\,\ref{fig:transport}\,(c) we extract an upper critical field of $B_\mathrm{c2,ip} = \SI{2.84\pm0.05}{T}$ for an in-plane (ip) geometry and $B_\mathrm{c2,oop} = \SI{2.04\pm0.02}{T}$ for out-of-plane (oop). Such high critical fields enable the use of air bridges in applications where magnetic fields up to $\SI{2}{T}$ are needed, such as electron spin resonance spectroscopy \cite{bienfait.2016, wang.2023}, superconducting qubit experiments \cite{krause.2022}, or parametric amplifiers that operate at high fields \cite{janssen.2024}.

\begin{figure}
\includegraphics{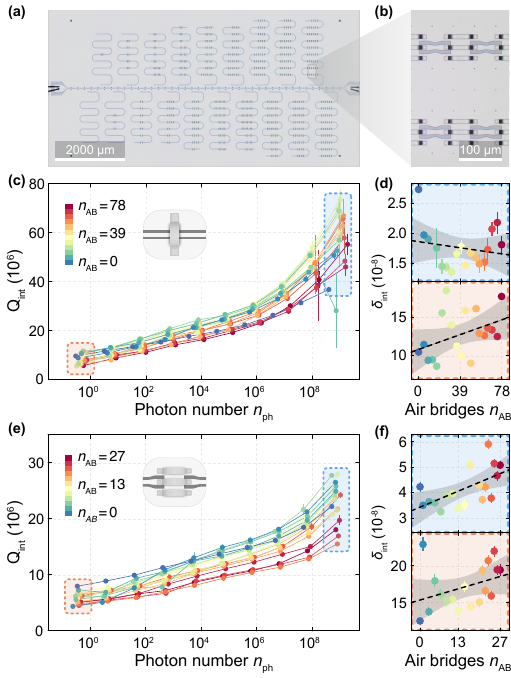}
\caption{\label{fig:resonator-q}CPW resonator measurements with different air bridge configurations. \textbf{(a)} Optical microscope image of 19 individual $\lambda/4$ resonators coupled to a common feedline. \textbf{(b)} Magnified view of in-line air bridges in the signal line of a CPW. \textbf{(c)} Internal quality factor $Q_\mathrm{int}$ as a function of photon number for 19 different resonators on a single chip. The number of grounding air bridges is varied between $n_\mathrm{AB}=0$ and $n_\mathrm{AB}=78$. \textbf{(d)} Internal loss $\delta_\mathrm{int}$ versus $n_\mathrm{AB}$ in the high-power (blue) and low-power (orange) regime for the grounding configuration. As visualized by the gray $\SI{95}{\percent}$ confidence interval, no clear dependence is visible. \textbf{(e)} Internal quality factor as a function of photon number for 18 different resonators on a single chip. Up to 27 in-line air bridges serve as interconnects of the CPW segments. \textbf{(f)} Loss $\delta_\mathrm{int}$ versus $n_\mathrm{AB}$ in the high- and low-power regime in the in-line configuration. In analogy to (d), no clear dependence on the number of air bridges is visible in the single-photon limit.}
\end{figure}

In order to analyze dielectric properties of the air bridges, we probe different CPW resonator configurations and extract the internal quality factor $Q_\mathrm{int}$ \cite{McRae.2020}. We measure the complex-valued scattering parameter $S_{21}$ as a function of frequency around the fundamental resonance frequency of each resonator with a vector network analyzer. The data is fitted by a circle fit model, which allows us to extract $Q_\mathrm{int}$ at different photon numbers \cite{Probst.2015}. This figure of merit defines the internal loss in our system and can be in particular used to extract the additional loss per air bridge \cite{chen2014} by probing resonators with varying numbers of air bridges. 
The chips are wire-bonded after the chemical treatment and mounted in a dilution refrigerator system operating at a base temperature of $\SI{10}{mK}$. We control the number of photons in the CPW resonators by varying the drive power of our signal between $\SI{-70}{dBm}$ and $\SI{-160}{dBm}$ at the sample. The transmitted signal is then amplified by a low-noise high electron mobility transistor (HEMT) and an additional room temperature amplifier. The measurement setup is explained in more detail in Appendix \ref{sec:supp-setup}. We focus on two different resonator configurations: first the \textit{grounding} configuration, in which we place the air bridge as an additional grounding element across the CPW, and second, the \textit{in-line} configuration, in which we use the air bridge as part of the center conductor of the CPW. Microscope images of the latter device are shown in Figs.\,\ref{fig:resonator-q}\,(a) and (b). 
All resonators are impedance-matched to the $\SI{50}{\Omega}$ microwave environment with a CPW center conductor width of $w=\SI{10}{\micro m}$ and a gap of $g=\SI{6}{\micro m}$. The $\lambda/4$ resonators are coupled to the feedline with a coupling quality factor of roughly $\SI{4e6}{}$. 
For both \textit{grounding} air bridges and \textit{in-line} air bridges, we measure a reduction in the internal quality factor when reducing the input power and therefore the effective photon number in the resonator (cf.\ Fig.\,\ref{fig:resonator-q}\,(c) and Fig.\,\ref{fig:resonator-q}\,(e), respectively). 
This is attributed to the TLS loss, as typically observed in superconducting qubits and resonators \cite{Phillips.1987, Muller.2019}.

The \textit{grounding} air bridge chip features 19 resonators with fundamental resonances between $\SI{4}{GHz}$ and $\SI{6}{GHz}$ and a number of air bridges varying between $n_\mathrm{AB}=0$ and $n_\mathrm{AB}=78$. Each grounding air bridge has a width of \SI{20}{\micro m} and a length of \SI{60}{\micro m}.
In the single-photon regime we reach a mean quality factor of $Q_\mathrm{int}=\SI{8.23\pm1.84e6}{}$, with the best-performing resonator surpassing $\SI{11.5\pm0.4e6}{}$, containing 17 grounding air bridges. 
We analyze the internal loss $\delta_\mathrm{int} \approx \tan\delta_\mathrm{int} = 1/Q_\mathrm{int}$ as a function of air bridge number $n_\mathrm{AB}$ in the high power (blue) and low power (orange) regime. Using a linear regression model with a \SI{95}{\percent} confidence interval, we fit the data points in the blue shaded region in Fig.\,\ref{fig:resonator-q}\,(d). We exclude the green resonance from the fit, since the extracted internal quality factor has a large uncertainty due to a nonlinear response. From the fit, we obtain a negative loss per air bridge in the high power limit of $\SI{-2.58\pm7.11e-11}{}$ with a resulting coefficient of determination of $R^2=0.04$. This clearly shows that the loss per air bridge is below the resolution limit of our resonators.
Additionally, we extract the loss per air bridge in the single-photon limit in the orange shaded region in Fig.\,\ref{fig:resonator-q}\,(d). The extracted loss per air bridge of $\SI{4.92\pm5.67e-10}{}$ is an order of magnitude lower compared to the best reported air bridge fabrication methods \cite{bu.2025}. However, as in the high power regime, the resulting coefficient of determination $R^2=0.16$ still indicates that the loss per air bridge is below the resolution limit of the sample. The large uncertainty in the loss estimate arises from the fact that the dissipation per bridge is less than the resonator-to-resonator variation in loss.

We perform an identical analysis for the \textit{in-line} chip with 19 individual resonators that contain air bridges ranging between $n_\mathrm{AB}=0$ and $n_\mathrm{AB}=27$. Each in-line air bridge is fabricated with a center conductor width of \SI{10}{\micro m} and a length of \SI{60}{\micro m}. In order to imitate CPW crossovers with grounding, each in-line air bridge has two identical passive neighboring air bridges, as shown in Fig.\,\ref{fig:resonator-q}\,(b). One resonance is excluded due to a fabrication defect, likely caused by dust in the coplanar waveguide.
The resulting quality factors are visualized as a function of photon number in Fig.\,\ref{fig:resonator-q}\,(e). We obtain a mean quality factor of $Q_\mathrm{int}=\SI{6.00\pm0.98e6}{}$ in the single-photon regime. While the overall quality factor is lower both in the single-photon and the high-power regime compared to the grounding air bridge chip, we attribute this behaviour to wafer-to-wafer variations in resonator performance, and not to the air bridge fabrication process itself. This is confirmed by an analysis of the loss per air bridge equivalent to the analysis above.
We extract the loss per air bridge in the high power regime from the blue shaded region in Fig.\,\ref{fig:resonator-q}\,(f). The obtained loss per air bridge is \SI{4.89\pm3.46e-10}{} with $R^2=0.36$. In the single-photon regime (orange), we extract a loss per air bridge of \SI{1.18\pm1.60e-9}{} ($R^2=0.13$), which is notably larger compared to the loss per grounding air bridge. While a larger loss is in principle expected due to a higher participation ratio for in-line air bridges, this increase mainly attributed to the lower number of air bridges compared to the grounding resonator sample. By probing 27 instead of 78 bridges, the resolution limit of our resonators is lower, which is reflected in the larger uncertainty.

\section{\label{sec:bruckmon}Vacuum Gap Air Bridge Qubit}

\begin{figure}
\includegraphics{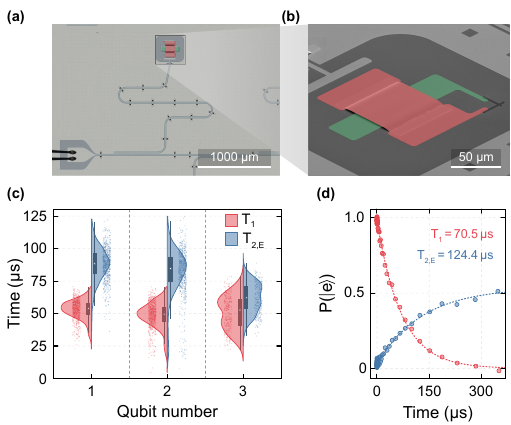}
\caption{\label{fig:qubit}Transmon qubit with a vacuum gap capacitance. \textbf{(a)} Optical microscope image of a vacuum-gap air bridge qubit. The bottom island is highlighted in green, the top island including the bridge in red. \textbf{(b)} SEM image of a fabricated air bridge qubit. The Josephson junction is highlighted in black. \textbf{(c)} Time domain measurements of three air bridge qubits, averaged over 100 hours. The left side of the violin plot (red) visualizes the histogram of the lifetime $T_1$, whereas the right side (blue) visualizes the echo coherence time $T_\mathrm{2,E}$. Individual traces are visualized as single points in the corresponding color. \textbf{(d)} Single trace of a qubit lifetime $T_1$ and coherence time $T_\mathrm{2,E}$ from qubit two.}
\end{figure}

As the Nb air bridge fabrication process yields stable metallization layers for sizes up to $150\times\SI{150}{\micro m\squared}$, we can utilize air bridges as versatile vacuum-gap capacitors \cite{cicak.2010}. In particular, we use air bridges as a shunting capacitance of transmon qubits by vertically stacking the capacitor pads. This pushes the energy stored in the electric field from the substrate into loss-free vacuum. Furthermore, because of the larger capacitance per area the overall footprint of qubits is reduced, similar to designs that utilize flip-chip geometry \cite{li.2021}, through-silicon vias \cite{hazard.2023a}, or suspended silicon beams \cite{zemlicka.2023}.
In our design, the bottom capacitor plate has dimensions of $\SI{240}{\micro m}\times\SI{60}{\micro m}$, and the air bridge on top has a width of $\SI{120}{\micro m}$ and a length of $\SI{105}{\micro m}$. The height of the bridge is $\SI{2.7}{\micro m}$. An optical microscope image as well as an SEM image is shown in Fig.\,\ref{fig:qubit}\,(a) and (b). We simulate the resulting structure with a finite element model (FEM) solver to extract the electric field and capacitances.
Compared to roughly $\SI{10}{\percent}$ in conventional coplanar transmons, $\SI{42}{\percent}$ of the energy stored in the electric field resides in vacuum. Moreover, $\SI{86}{\percent}$ of the field energy stored in vacuum is located between the capacitor plates.

We fabricate a sample with multiple air bridge qubits according to the fabrication steps presented in Section \ref{sec:fabrication}. In order to fabricate the Josephson junction, we use the Manhattan shadow evaporation process \cite{Costache.2012, Kreikebaum.2020} with a CSAR 6200 bottom resist and a PMMA 950k top resist mask. The bottom and top Al electrodes have thicknesses of $\SI{30}{nm}$ and $\SI{70}{nm}$, respectively. After junction fabrication, we galvanically connect the Al with the Nb groundplane using an ex-situ Al bandaging process \cite{Dunsworth.2017}. The junction fabrication process is explained in more detail in Appendix\,\ref{sec:supp-qubit-fab}.
The air bridge qubits have transition frequencies between $\SI{4.3}{GHz}$ and $\SI{5.3}{GHz}$ and a charging energy $E_\mathrm{C}$ between $\SI{240}{MHz}$ and $\SI{245}{MHz}$. They are dispersively coupled to readout resonators with fundamental frequencies between $\SI{6.9}{GHz}$ and $\SI{7.4}{GHz}$ and a coupling strength of $\SI{60}{MHz}$ between resonator and qubit. In order to acquire statistically relevant data, we measure three air bridge qubits on one sample over 100 hours in two cooldowns. The resulting data is visualized in Fig.\,\ref{fig:qubit}\,(c) as a violin plot.
The median lifetimes $T_1$ (red) for all three qubits are in the range of $\SI{49.4}{\micro s}$ to $\SI{53.5}{\micro s}$, with single traces exceeding $\SI{75}{\micro s}$. Exemplary single traces are shown in Fig.\,\ref{fig:qubit}\,(d).
Median echo coherence times $T_\mathrm{2,E}$ (blue) are $\SI{88.3}{\micro s}$ and $\SI{84.3}{\micro s}$ for qubit one and two, respectively. The coherence time $T_\mathrm{2,E} = \SI{62}{\micro s}$ of qubit three is lower, since it has been measured in a dilution refrigerator setup with less filters.
The median $T_1$ time is reduced by a factor of 2 to 5 compared to state-of-the-art coplanar Nb transmons \cite{bal.2024, tuokkola.2024}. We attribute this to the higher metal-air interface participation ratio in the vacuum gap air bridge qubit, since amorphous Nb oxides are known to be a major loss channel in transmon qubits \cite{bal.2024}.

\section{\label{sec:summary}Conclusion and Outlook}

In conclusion, we have demonstrated a novel air bridge fabrication process based on a patterned hard mask, which is highly scalable and enables high yield due to its subtractive processing approach. By incorporating multiple cleaning steps, we show that the extracted single-photon loss per grounding air bridge of $\delta_\mathrm{AB} = \SI{4.92\pm5.67e-10}{}$ and per in-line air bridge of $\delta_\mathrm{AB} = \SI{1.18\pm1.60e-9}{}$ remain below the resolution limit of the resonator-based measurement. Furthermore, we have verified that the fabricated air bridges support superconductivity up to \SI{9.3}{K} at zero magnetic field and up to \SI{3.9}{K} in a magnetic field of \SI{1.6}{T}. This extends their applicability to high magnetic fields and temperatures above \SI{4}{K}, which is not feasible with conventional aluminum-based approaches. 

An open question that has not been addressed is the long-term stability, as Nb resonators are known to oxidize and decrease in quality over time \cite{Verjauw.2021}. While we wanted to focus on the process itself within this study, we are planning on investigating the aging behavior, as well as the compability with surface treatments such as buffer and capping materials to avoid aging \cite{bal.2024}.

Beyond their use as interconnects, air bridges can also serve as active elements in tunable coupler qubits \cite{Warren.2023, bu.2025}. The demonstrated low-loss air bridges hold significant potential for reducing coupler-induced decoherence compared to existing fabrication techniques. Additionally, we have shown that large air bridges can function as capacitors in transmon qubits, yielding a median qubit lifetime of \(T_1 = \SI{51.6}{\micro s}\). While the dominant decoherence mechanisms require further investigation, this qubit design offers the potential for a smaller footprint compared to planar transmons by reducing the bridge height and increasing capacitance. Moreover, the confinement of the electric field between capacitor leads enhances robustness against misalignment in 3D-integrated qubit architectures.

Finally, since the resist is protected by the hard mask during milling and deposition, our process is compatible with a variety of other materials, including Ta and NbTiN, as well as any material that remains intact during the hard mask removal process. This versatility makes the proposed fabrication technique well-suited not only for superconducting circuits but also for a wide range of micro- and nanoelectronic devices.

\begin{acknowledgments}
This work received financial support from the German Federal Ministry of Education and Research via the funding program ’Quantum technologies - from basic research to the market’ under contract number 13N15680 (GeQCoS) and under contract number 13N16188 (MUNIQC-SC), by the Deutsche Forschungsgemeinschaft (DFG, German Research Foundation) via project number FI2549/1-1 and via the Germany’s Excellence Strategy EXC- 2111-390814868 ‘MCQST’ as well as by the European Union by the EU Flagship on Quantum Technology HORIZON-CL4-2022-QUANTUM-01-SGA project 101113946 OpenSuperQPlus100. The research is part of the Munich Quantum Valley, which is supported by the Bavarian state government with funds from the Hightech Agenda Bayern Plus.
\end{acknowledgments}

\section*{Data Availability}
The data that support the findings of this paper are openly available~\cite{Bruckmoser.Zenodo}.

\appendix

\begin{figure}
\includegraphics[width=0.47\textwidth]{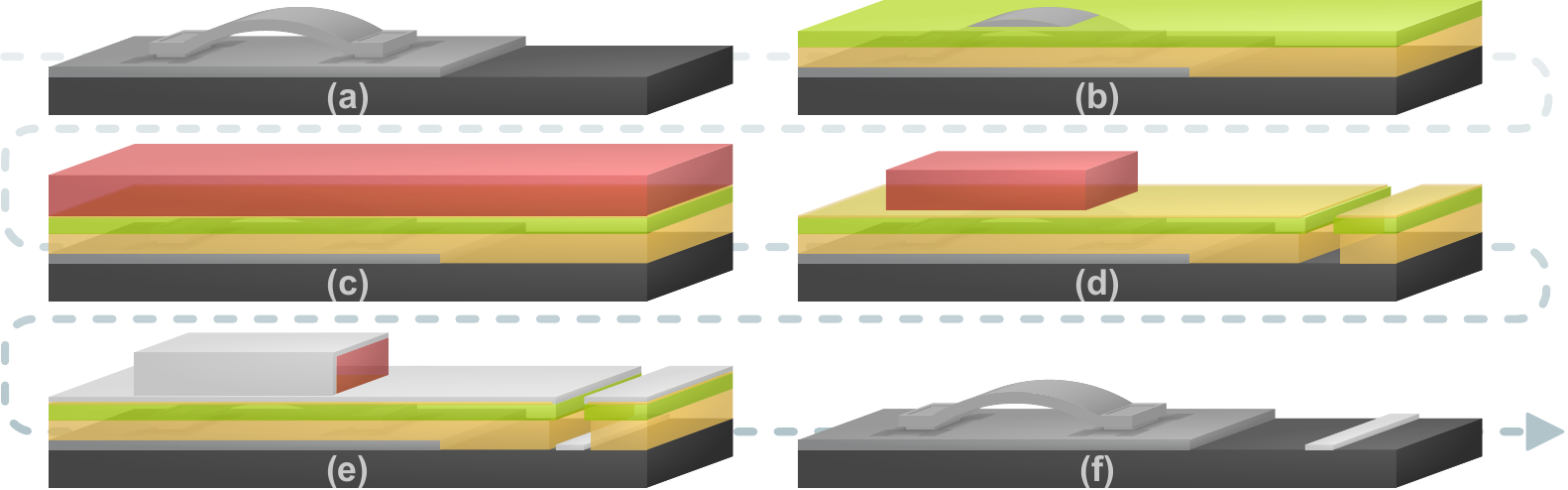}
\caption{\label{fig:fab-supp}Fabrication steps to protect air bridges during Josephson junction fabrication, not to scale. (a) Chip with patterned groundplane and finished air bridges. (b) Spin coating of bilayer electron beam lithography resist. (c) Coating of additional bilayer consisting of a sacrificial CSAR layer and a thick optical resist. (d) Subsequent exposure and development of optical resist and electron beam resist. (e) Double angle shadow evaporation of Al to form Josephson junctions. (f) tear-off free lift-off.}
\end{figure}

\section{\label{sec:supp-qubit-fab}Qubit Fabrication}

In order to fabricate Josephson junctions, we use the well-established Manhattan shadow evaporation process. However, to realize Josephson junction sizes in the order of $100\times \SI{100}{nm\squared}$, the total electron beam lithography (EBL) resist height does not exceed $\SI{1}{\micro m}$ for sufficiently large aspect ratios. The air bridge is thus higher than the resist and needs to be protected with an additional optical lithography step to prevent lift-off tear-offs of Al at the air bridge interface. 

The fabrication steps are visualized in Fig.\,\ref{fig:fab-supp}. We start the Josephson junction fabrication process by spin coating a $\SI{650}{nm}$ thick bottom layer of CSAR 6200, followed by a less sensitive $\SI{350}{nm}$ thick top layer of PMMA 950K. In order to exploit complementary exposure techniques and developers, we coat the top layer with an additional $\SI{50}{nm}$ of CSAR 6200. As a last step, we coat the chip with a thick layer of optically sensitive resist mr-DWL 5, which is selectively removed around air bridges in a solvent based developer (mr-Dev 600). The thin sacrificial CSAR layer serves to protect the underlying PMMA from being attacked by the optical developer. Afterwards, we expose and develop the EBL resist and evaporate Al at a tilt of \SI{35}{\degree} with a rotation of \SI{0}{\degree} and \SI{90}{\degree} to fabricate Josephson junctions. During deposition, the air bridges are fully protected by the resist, which prevents Al tear-offs in the lift-off process step.

In a subsequent step we repeat this process to fabricate ex-situ bandages with an additional milling step prior to Al deposition.

\section{\label{sec:supp-stress}Niobium film stress}

\begin{figure}
\vspace{10pt}
\includegraphics[width=0.47\textwidth]{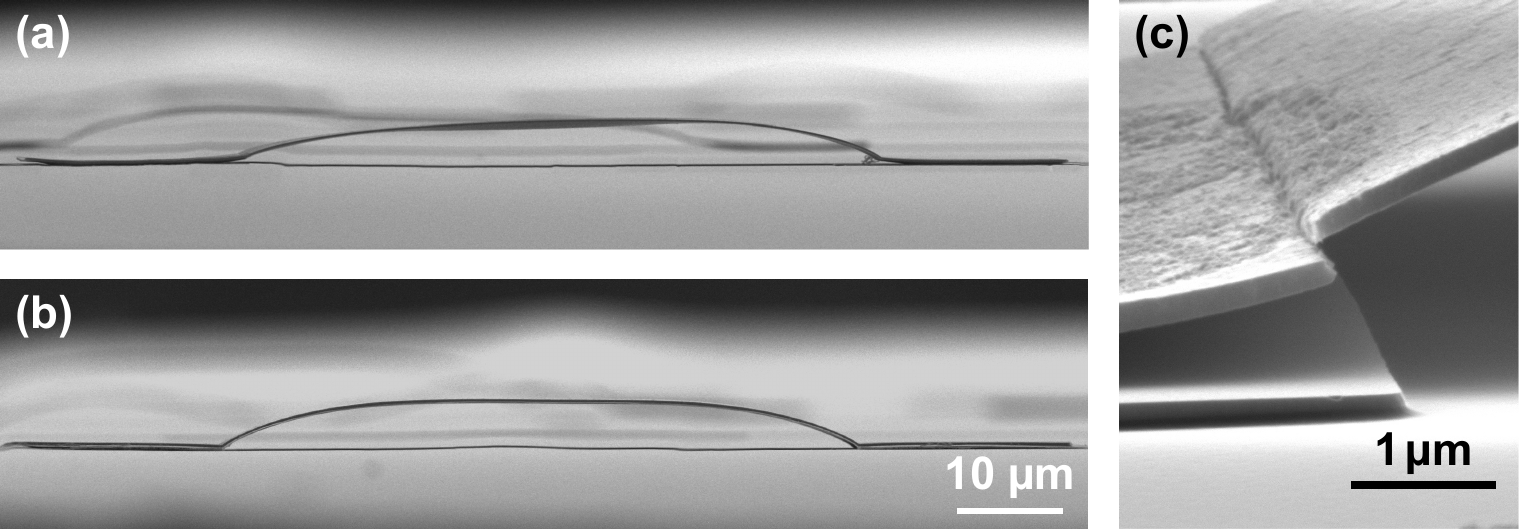}
\caption{\label{fig:stress}SEM image of air bridges sputtered at different process conditions. (a) Air bridge sputtered at a partial pressure of $\SI{3.3}{mTorr}$. The center is bent due to tensile stress in the film. (b) Air bridge sputtered at a partial pressure of $\SI{2.0}{mTorr}$. (c) Broken connection between bridge and pad for a tensile stressed air bridge.}
\end{figure}

To ensure a high yield of air bridges, it was essential for us to optimize the Nb sputtering conditions. Our typical sputtering regime utilizes an argon partial pressure of $\SI{3.3}{mTorr}$. However, the air bridge metallization layer bends due to the high intrinsic stress that results from the sputtering conditions \cite{imamura.1989}. This is clearly visible in the center part of the \SI{300}{nm} thick air bridge in Fig.\,\ref{fig:stress}\,(a). In particular, this can lead to breaking points and ultimately to collapses (Fig.\,\ref{fig:stress}\,(c)). At an even higher partial pressure, the film peels off after deposition.
By reducing the partial pressure, we were able to reduce tensile stress in our film. An exemplary relaxed air bridge sputtered at a lower pressure of $\SI{2.0}{mTorr}$ is shown in Fig.\,\ref{fig:stress}\,(b). We hope to quantify the intrinsic stress and further optimize the yield for higher bridge lengths by employing local asymmetric and Bragg-Brentano x-ray diffraction configurations \cite{motazedian.2019}.

\section{\label{sec:supp-setup}Measurement setup}
\begin{figure}
\includegraphics{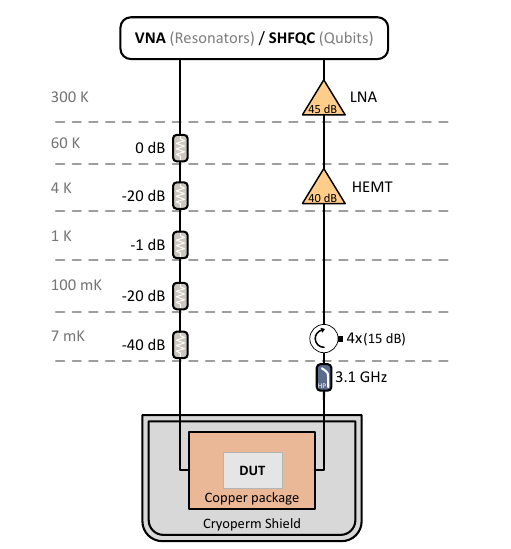}
\caption{\label{fig:wiring}Measurement setup for both resonator and qubit devices. For details, see main text.}
\end{figure}

\begin{figure}
\includegraphics{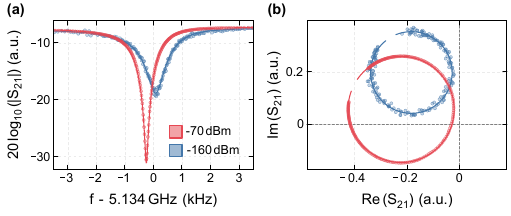}
\caption{\label{fig:traces}Exemplary measurement trace of a resonator containing 31 grounding air bridges. (a) Magnitude of $S_{21}$ plotted against the frequency for both single photons (blue) and $10^9$ photons (red). Circles correspond to measurement points, the solid line corresponds to the fit. (b) The same traces, plotted in the complex plane.}
\end{figure}

The sample is mounted at the mixing chamber stage of a dilution refrigerator (Bluefors Fast Sample Exchange cryostat or Qinu Sionludi L). Fig.\,\ref{fig:wiring} shows the electronic wiring from room temperature up to Millikelvin. We use two different devices for measurements: resonators are probed with a Keysight vector network analyzer (VNA), qubits are being controlled and read out with a Zurich Instruments quantum controller (SHFQC). We attenuate the input signal by a total of \SI{81}{dB}. In addition, we characterized the cable loss in the input line, which results in an additional \SI{8}{dB} of loss at \SI{4}{GHz}. We take this into account for all frequencies when determining the estimated photon number in the $\lambda/4$ resonators.
For measurements in the Bluefors cryostat, the output signal is routed through a switch, the Qinu cryostat does not contain a switch. Two dual-junction isolators ensure uni-directional signal routing. An additional high-pass filter (Minicircuits VHF-3100+) protects the sample from noise. We amplify the transmitted signal with a HEMT amplifier at the 4K stage of the cryostat, and an additional low-noise amplifier (LNA) at room temperature.
The sample is wire-bonded to a gold-coated PCB and placed inside a copper package. The package is located inside two cryoperm shields to protect the sample from stray magnetic fields.
An exemplary resonator response at $\SI{5.1}{GHz}$ with 31 grounding air bridges is shown in Fig.\,\ref{fig:traces}. The red trace is measured at a drive power of $\SI{-70}{dBm}$, corresponding to roughly $10^9$ photons. The blue trace is measured at a drive power of $\SI{-160}{dBm}$, which resembles roughly one photon. The single photon quality factor of this trace corresponds to $Q_\mathrm{int} = \SI{11.05\pm0.28e6}{}$. For a drive power of $\SI{-70}{dBm}$, we fit an internal quality factor of $Q_\mathrm{int}=\SI{59.5\pm2.6e6}{}$. Although the signal-to-noise-ratio is significantly better, the uncertainty in $Q_\mathrm{int}$ is larger due to undercoupling, arising from the constant coupling quality factor of $Q_\mathrm{c}=\SI{4.1e6}{}$. We try to achieve critical coupling of our resonators in the low-power regime, as this is the main region of interest. Also, the strong drive power starts to result in a slight nonlinearity, which we attribute to the saturation limit of our HEMT amplifier.

\section{\label{sec:supp-dc}DC conductivity}
\begin{figure}
\includegraphics{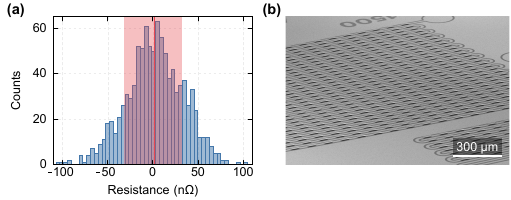}
\caption{\label{fig:conductivity}DC conductivity measurements of 1500 air bridges in series at $\SI{3}{K}$ to extract the resistance of the metal-metal interface between ground plane and air bridge. (a) Histogram of 1000 measurement shots, resulting in a mean total resistance value of $R_\mathrm{tot}=\SI{2}{n\ohm}$ and a standard deviation of $\sigma=\SI{32}{n\ohm}$. (b) SEM image of the measured device with a chain of 1500 air bridges.}
\end{figure}

In order to investigate the resistivity of the metal-metal interface between ground plane and air bridge, we fabricated a daisy chain consisting of 1500 air bridges in series, each with a pad size of $\SI{15}{\micro m}\times\SI{15}{\micro m}$, resulting in a total interface area of $1500\times2\times\SI{15}{\micro m}\times \SI{15}{\micro m} =\SI{0.675}{\mm\squared}$. The histogram of 1000 measurement shots measured at $\SI{3}{K}$ with a current of $\SI{100}{mA}$ is shown in Fig.\,\ref{fig:conductivity}. We measure a resistance of $R_\mathrm{tot}=\SI{2\pm32}{n\ohm}$ at $\SI{3}{K}$, corresponding to a contact resistance of $R=\SI{3\pm47}{f\ohm}$ per $\SI{}{\micro m\squared}$. This confirms a low-ohmic galvanic connection between ground plane and air bridge. 

\newpage

\bibliography{airbridge-lib}

\end{document}